\pdfoutput=1
\documentclass[sigconf]{acmart}

\AtBeginDocument{%
  \providecommand\BibTeX{{%
    \normalfont B\kern-0.5em{\scshape i\kern-0.25em b}\kern-0.8em\TeX}}}

\setcopyright{acmcopyright}
\copyrightyear{2024}
\acmYear{2024}
\setcopyright{acmlicensed}\acmConference[WWW '24]{Proceedings of the ACM Web Conference 2024}{May 13--17, 2024}{Singapore, Singapore}
\acmBooktitle{Proceedings of the ACM Web Conference 2024 (WWW '24), May 13--17, 2024, Singapore, Singapore}
\acmDOI{10.1145/3589334.3648158}
\acmISBN{979-8-4007-0171-9/24/05}

\usepackage{enumitem}
\usepackage{algorithmic}
\usepackage[ruled]{algorithm2e}
\usepackage{graphicx}
\usepackage{textcomp}
\usepackage{xcolor}
\usepackage{bm}
\usepackage{balance} 
\usepackage{multirow}
\usepackage{ragged2e}
\usepackage{caption}
\usepackage{subcaption}
\usepackage[normalem]{ulem}
\useunder{\uline}{\ul}{}
\usepackage{amsmath}
\usepackage{amsthm}
\usepackage{setspace}
\usepackage{epstopdf}
\usepackage{colortbl}

\newcommand{\ie}{\emph{i.e., }}
\newcommand{\eg}{\emph{e.g., }}

\newcommand{\wrt}{\emph{w.r.t. }}

\begin{document}

\title{Item-side Fairness of Large Language Model-based Recommendation System}

\settopmatter{authorsperrow=4}
\author{Meng Jiang}
\affiliation{%
  \institution{University of Science and Technology of China}
  \streetaddress{443 Huangshan Rd}
  \city{Hefei}
  \country{China}}
\email{jiangm@mail.ustc.edu.cn}

\author{Keqin Bao}
\affiliation{%
  \institution{University of Science and Technology of China}
  \streetaddress{443 Huangshan Rd}
  \city{Hefei}
  \country{China}}
\email{baokq@mail.ustc.edu.cn}

\author{Jizhi Zhang}
\affiliation{%
  \institution{University of Science and Technology of China}
  \streetaddress{443 Huangshan Rd}
  \city{Hefei}
  \country{China}}
\email{cdzhangjizhi@mail.ustc.edu.cn}

\author{Wenjie Wang}
\affiliation{%
  \institution{National University of Singapore}
  \country{Singapore}}
\email{wenjiewang96@gmail.com}
\authornote{Corresponding authors.}

\author{Zhengyi Yang}
\affiliation{%
  \institution{University of Science and Technology of China}
  \streetaddress{443 Huangshan Rd}
  \city{Hefei}
  \country{China}}
\email{yangzhy@mail.ustc.edu.cn}

\author{Fuli Feng}
\affiliation{%
  \institution{University of Science and Technology of China}
  \streetaddress{443 Huangshan Rd}
  \city{Hefei}
  \country{China}}
\email{fulifeng93@gmail.com}

\author{Xiangnan He}
\affiliation{%
  \institution{MoE Key Lab of BIPC}
  \streetaddress{443 Huangshan Rd}
  \city{Hefei}
  \country{China}}
\email{xiangnanhe@gmail.com}
\authornotemark[1]








\renewcommand{\shortauthors}{Meng Jiang, et al.}

\begin{abstract}

Recommendation systems for Web content distribution intricately connect to the information access and exposure opportunities for vulnerable populations. The emergence of Large Language Models-based Recommendation System (LRS) may introduce additional societal challenges to recommendation systems due to the inherent biases in Large Language Models (LLMs). From the perspective of item-side fairness, there remains a lack of comprehensive investigation into the item-side fairness of LRS given the unique characteristics of LRS compared to conventional recommendation systems. To bridge this gap, this study examines the property of LRS with respect to item-side fairness and reveals the influencing factors of both historical users' interactions and inherent semantic biases of LLMs, shedding light on the need to extend conventional item-side fairness methods for LRS. Towards this goal, we develop a concise and effective framework called IFairLRS to enhance the item-side fairness of an LRS. IFairLRS covers the main stages of building an LRS with specifically adapted strategies to calibrate the recommendations of LRS. We utilize IFairLRS to fine-tune LLaMA, a representative LLM, on \textit{MovieLens} and \textit{Steam} datasets, and observe significant item-side fairness improvements. The code can be found in \href{https://github.com/JiangM-C/IFairLRS.git}{https://github.com/JiangM-C/IFairLRS.git}.

\end{abstract}






\maketitle

\section{Introduction}


Recommendation systems play a pivotal role in the distribution of diverse Web content, encompassing news reports~\cite{newsReport}, social posts~\cite{S_posts}, micro-videos~\cite{videos}, and a range of descriptions like clothing~\cite{clothing}, cuisine~\cite{cuisine}, pharmaceuticals~\cite{pharmaceuticals}, and job~\cite{job}. Recommendation systems are intricately linked to societal concerns such as equitable information access and exposure opportunities for vulnerable populations. Recently, Large Language Models (LLMs)~\cite{LLaMA,GPT} have gained prominence as a key in recommendation systems due to their exceptional content comprehension capabilities~\cite{tallrec,bigrec,instructrec}. 
Nevertheless, the integration of LLMs may introduce societal challenges to recommendation systems, primarily stemming from the inherent biases in LLM training datasets~\cite{GPT4, LLaMA}.  
Therefore, there is a pressing need to investigate the trustworthiness of LLM-based Recommendation Systems (LRS).

Item-side Fairness (IF)~\cite{yongfeng_survey, minzhang_survey} is a critical aspect of trustworthy recommendations, aiming to provide fair exposure opportunities for different item groups. IF is widely used to ensure the rights and profit of item producers such as the opportunity to seek appropriate candidates of micro-businesses in job recommendation~\cite{job2}. Additionally, IF can also be applied to different topics, facilitating effective dissemination of content (\eg news reports) on societal issues like environmental sustainability and climate action. Furthermore, IF can elevate the visibility of content related to vulnerable populations, ensuring their interests and demands receive adequate attention from society. 
Despite its significance, 
there is a lack of comprehensive research investigating IF in the context of LLM-based recommendation systems.


LLM-based recommendation systems~\cite{a1,instructrec,open_world_rec} exhibit unique characteristics compared to conventional recommendation systems~\cite{llm_rec_survey_1, llm_rec_survey_3}. These include their reliance on semantic clues to infer user preferences, the use of instructions to describe the recommendation task, and the generation of recommendations instead of relying solely on
discriminative predictions. 
Consequently, previous findings regarding item-side fairness in conventional methods may not hold true for LLM-based systems. Therefore, it is crucial to examine the property of LRS with respect to item-side fairness. This study specifically investigates two key questions:
\begin{itemize}[leftmargin=*]
    \item \textbf{RQ1:} How does LRS perform in terms of item-side fairness compared to traditional recommendation models?
    \item \textbf{RQ2:} What is the root cause of the fairness issue in the
LRS?
    %
\end{itemize}

To answer these questions, we conduct exploratory experiments on two public datasets under a sequential recommendation setting as per recent LRS studies~\cite{bigrec,tallrec}. We compare LLM-based methods, represented by BIGRec~\cite{bigrec}, with SASRec~\cite{SASRec}, representative of conventional sequential recommendation methods in terms of item-side fairness. Our findings indicate that LRS is notably influenced by the popularity factor, as BIGRec consistently recommends more popular items compared to SASRec. Additionally, BIGRec exhibits biases towards certain item groups in specific genres (\eg crime), suggesting the impact of inherent semantic biases within the LLM since item genre is related to the textual description of items (inputs of LRS). In summary, the imbalanced distribution of historical interactions for LRS training and the inherent semantic bias are both contributing factors to the unfairness observed in LRS. Consequently, it is crucial to adapt conventional methods to enhance item fairness by addressing these factors in LRS.

To achieve this target, we develop IFairLRS, a concise and effective framework that enhances the item-side fairness of an LRS. 
IFairLRS calibrates the recommendations of LRS to meet the IF requirements by considering the two stages of building an LRS: in-learning, and post-learning. 
For each stage, IFairLRS incorporates specific strategies adapted from IF methods for conventional recommendation models~\cite{ips, PDA}. 
In the in-learning stage, IFairLRS reweights training samples based on the bias observed between the distribution of target items and historical interactions. 
In the post-learning stage, IFairLRS can rerank the recommendations of LRS by incorporating a punishment term regarding unfairness. 
Experiments on real-world datasets demonstrate the effectiveness of our framework in enhancing item-side fairness \wrt both historical interactions and semantic biases. Our codes and data will be released upon acceptance.

In conclusion, our contributions are as follows:
\begin{itemize}[leftmargin=*]




\item We reveal the unfairness issue of LRS and the unique characteristics of LRS in terms of item-side fairness.

\item We develop a concise and effective framework for the item-side fairness of LRS with strategies for different training phases.

\item We conduct extensive experiments on real-world datasets, validating the effectiveness of the proposed framework with an in-depth analysis of the pros and cons of different strategies.
\end{itemize}
\section{Related Work}
In this section, we review the recent work on item-side fairness in recommendation and LLM-based recommendation systems.
\subsection{Item-side Fairness in Recommendation}
The item-side fairness considers whether the items are treated fairly by the recommendation system~\cite{yongfeng_survey, minzhang_survey}, and can be categorized into two primary classes: individual fairness and group fairness. Individual fairness requests each individual item be treated similarly~\cite{item_individual_1, item_individual_2, item_individual_3, item_individual_4} and group fairness demands each predefined item group be treated equally~\cite{item_group_1, item_group_2, item_group_4, item_group_5, item_group_6}. 
Our work falls under the category of group fairness. Different from the previous work, we examine the IF of LRS.

To the best of our knowledge, there is only one work LLMRank~\cite{LLMRank} associated with IF in LRS. 
However, LLMRank merely qualitatively observes the correlation between the frequency of item appearances in the recommendation results of LRS and the popularity of the item, without further systematic analysis like the quantitative study.
Different from LLMRank, we design multiple metrics to quantitatively analysis the IF in LRS from popularity and genre two perspectives and systematically propose methods to improve the IF of LRS.

\subsection{LLM-based Recommendation System}
LLMs, which consist of billions of parameters, have reformulated the paradigm of natural language processing~\cite {COT, llm_reasoner}. 
Different from traditional language models like BERT~\cite{bert}, GPT-2~\cite{gpt2}, BART~\cite{bart}, and so on, LLMs show much stronger natural language understanding and generation ability since LLMs have billions or more parameters, train on more data, and have more elaborate network structure~\cite{llm_survey, emergent, chatgpt, GPT4}. 
Among them, LLaMA~\cite{LLaMA} has garnered widespread acceptance in academia for its open-source nature. Thus, we adopt LLaMA as the LLM backbone of our research.

Astonished by the miracle exhibited by LLMs, there has sparked a trend in contemplating how to utilize the power of LLMs in the field of recommendation system~\cite{llm_rec_survey_2, open_world_rec, li2023exploring, a2, a3, a4, a5}. 
Current research on LRS follows a general pipeline: translating recommendation data into natural language input and then utilizing LLMs to generate recommendation results in a natural language form~\cite{ICL_1, LLMRank, fairllm, instructrec, rella}. 
However, due to limitations such as the lack of recommendation data during the pre-training phase of LLMs, directly using LLMs for recommendation can only achieve sub-optimal performance, making it necessary to tune LLMs on the recommendation data to unleash their recommendation capabilities~\cite{tallrec, bigrec, a6}. 
BIGRec~\cite{bigrec} has the potential to serve as a representative of the LRS approach since BIGRec encapsulates the fundamental elements of LRS and numerous methods can be conceptualized as further extensions grounded in the BIGRec paradigm.
Different from previous LRS user-side fairness evaluation work FaiRLLM~\cite{fairllm}, we take an exploratory approach from the item side, addressing a crucial aspect that is currently lacking in fairness research in LRS.


%

\section{Preliminary}
In this section, we first elaborate on the evaluation of item-side fairness in LLM-based recommendation systems.
Afterward, we introduce BIGRec \cite{bigrec}, one of the most recent LRS, which serves as the backbone model in our method.

\subsection{Evaluation of Item-side Fairness}
\label{sec:2_1}

Given a user with historical interactions, recommendation models usually rank all item candidates and return top-$K$ items as recommendations~\cite{dros}. To pursue fair recommendations across item groups, IF requires that the recommendation proportion of an item group $G$ should be calibrated to the proportion of $G$ in the users' historically liked items~\cite{Calibrated, Calibrated2}. 
In other words, recommendation models should neither over-recommend an item group nor decrease its recommendations compared to users' historical interactions. 
To ensure the protected group's rights and interests are not violated.

Formally, let $\mathcal{H}$ denote the set of all users' interaction sequences in the history, $\mathcal{P}$ denote the set of top-$K$ recommendations of all users at the inference phase, and $\mathcal{G}$ denote the set of item groups. Given an item group $G \in \mathcal{G}$, we can measure the recommendation proportion of group $G$ by 
\begin{equation}
\label{eq:count_predict}
    \textbf{GP}(G) = \frac{\sum_{P \in \mathcal{P}}\sum_{v \in P} \mathbb{I}(v \in G) }{\sum_{G' \in \mathcal{G}}\sum_{P \in \mathcal{P}}\sum_{v \in P} \mathbb{I}(v \in G')},
\end{equation}
where $\mathbb{I}(v \in G)$ is an identity function:
\begin{equation}
\mathbb{I}(v \in G) = 
\begin{cases}
1, & \text{item $v$ belongs to group $G$}  \\
0, & \text{otherwise} 
\end{cases}.
\end{equation}
Intuitively, $\textbf{GP}(G)$ calculates the recommendation proportion of group $G$ in the top-$K$ recommendations of all users. 

Accordingly, the interaction proportion of group $G$ in the historical interaction sequences $\mathcal{H}$ can be obtained by:
\begin{equation}
\label{eq:count}
    \textbf{GH}(G) = \frac{\sum_{H \in \mathcal{H}}\sum_{v \in H} \mathbb{I}(v \in G) }{\sum_{G' \in \mathcal{G}}\sum_{H \in \mathcal{H}}\sum_{v \in H} \mathbb{I}(v \in G')}.
\end{equation}
Thereafter, we can define \textbf{Group Unfairness (GU)} \wrt group $G$ to measure whether recommendation models amplify the recommendations of group $G$ or overlook its recommendations. Formally, 
\begin{equation}
\label{eq:GU}
    \textbf{GU}(G) = \textbf{GP}(G) - \textbf{GH}(G),
\end{equation}
where $\textbf{GU}(G) > 0$ suggests that the recommendation model tends to over-recommend items from group $G$, while $\textbf{GU}(G) < 0$ implies the recommendation model overlooks group $G$. Both reflect item-side unfairness at the group level.

Following prior studies \cite{Calibrated, Calibrated2}, we adopt two evaluation metrics to aggregate the item-side unfairness across groups. 
\begin{itemize}[leftmargin=*]
    \item \textbf{Mean Group Unfairness (MGU)} evaluates the overall fairness:
        \begin{equation} \label{eq:MGB}
            \textbf{MGU} = \frac{1}{|\mathcal{G}|}\sum_{G \in \mathcal{G}}|\textbf{GU}(G)|,
        \end{equation}
        \item \textbf{Disparity Group Unfairness (DGU)} measures the disparity between the maximum and minimum GU across the groups in $\mathcal{G}$, which evaluates the upper bound of item-side unfairness:
        \begin{equation} \label{eq:DGB}
            \textbf{DGU} = \max \{\textbf{GU}(G)\}_{G \in \mathcal{G}} - \min \{\textbf{GU}(G)\}_{G \in \mathcal{G}}.
        \end{equation}
\end{itemize}
In the experiments, we adopt the all-ranking protocol \cite{dros}, where we can vary $K$ to obtain top-$K$ recommendations so that we have $\textbf{GP}@K$, $\textbf{GU}@K$, $\textbf{MGU}@K$, and $\textbf{DGU}@K$, respectively.

\subsection{Brief on BIGRec}
\label{sec:bigrec}
Extensive research on LRS is emerging. To ensure the recommendation quality of LRS, substantial work has demonstrated that instruction-tuning is an indispensable phase~\cite{tallrec,bigrec}. Existing work often formulates recommendation data in natural language and tunes LLMs to generate personalized recommendations~\cite{bigrec, LLMRank, fairllm, instructrec}. In these studies, we select the representative BIGRec~\cite{bigrec} to assess IF in LRS. BIGRec is selected for its simplicity yet high effectiveness, and, more importantly, BIGRec embodies the fundamental elements of instruction-tuning-based LRS and many methods can be formulated as further extensions based on the paradigm of BIGRec. 


Following the general paradigm of applying LLMs in recommendation, BIGRec first represents user-item interaction data in natural language and employs instruction-tuning to fine-tune LLMs. In the inference stage, BIGRec generates item descriptions (\eg titles) as recommendations. 
However, the generated item descriptions might not exactly match any existing item. 
To deal with this issue, BIGRec designs a simple grounding paradigm, which leverages L2 embedding distance to ground generated item descriptions to existing items. 
Specifically, given \textbf{oracle} and $\textbf{emb}_i$ denoting the token embeddings (\ie latent representations) of the generated descriptions and the description of item $i$, respectively, BIGRec calculates their L2 distance by:
\begin{equation}
\label{eq:distance}
    D_i = ||\textbf{emb}_i - \textbf{oracle}||_2.
\end{equation}
Thereafter, BIGRec ranks items based on this L2 distance and returns the $K$-nearest items as the recommendations. 
\section{Probe the Item-side Fairness of LRS}
\label{sec:4}
To investigate the issue of item-side fairness in LRS, we initiated a preliminary experiment in this section by addressing the following research questions.

\begin{itemize}[leftmargin=*]
    \item \textbf{RQ1:} How does LRS perform in terms of item-side fairness compared to traditional recommendation models?
    \item \textbf{RQ2:} What is the root cause of the fairness issue in the LRS?
    %
\end{itemize}

\subsection{Experiment Setting}
\subsubsection{Datasets}
\label{sec:4_1_1}




\begin{table*}[!t]
\caption{Performance comparison of IF for groups split by popularity. The best results are indicated in boldface. We conduct each experiment three times and the average results are reported.}
\vspace{-0.3cm}

\label{tab:popularity_result}
\begin{tabular}{clllllllll}
\hline
Dataset                      & Model  & \textbf{MGU}@1           & \textbf{MGU}@5           & \textbf{MGU}@10          & \textbf{MGU}@20          & \textbf{DGU}@1           & \textbf{DGU}@5           & \textbf{DGU}@10          & \textbf{DGU}@20          \\ \hline
\multirow{2}{*}{MovieLens1M} & SASRec & \textbf{0.0238} & \textbf{0.0279} & 0.0287          & 0.0279          & \textbf{0.0846} & 0.0959          & 0.0976          & 0.0936          \\
                             & BIGRec & 0.1044          & 0.0285          & \textbf{0.0178} & \textbf{0.0128} & 0.3917          & \textbf{0.0956} & \textbf{0.0490} & \textbf{0.0415} \\ \hline
\multirow{2}{*}{Steam}       & SASRec & \textbf{0.0173} & \textbf{0.0166} & \textbf{0.0162} & \textbf{0.0157} & \textbf{0.0647} & \textbf{0.0485} & \textbf{0.0411} & \textbf{0.0445} \\
                             & BIGRec & 0.0411          & 0.0590          & 0.0736          & 0.0900          & 0.1586          & 0.1987          & 0.2460          & 0.3050          \\ \hline
\end{tabular}
\vspace{-0.4cm}

\end{table*}

We conducted experiments on two datasets containing information about different genres of items to perform an analysis of fairness regarding these genres.
\begin{itemize}[leftmargin=*]
    \item \textbf{MovieLens1M}~\cite{movie}\footnote{https://grouplens.org/datasets/movielens/1m/.} is a widely used dataset for a movie recommendation, which holds users' interaction records and movie genre. We retain titles as the textual information about films.
    \item \textbf{Steam}~\cite{steam}\footnote{https://jmcauley.ucsd.edu/data/amazon/.} provides user reviews for video games on the Steam platform. We adopt game titles as their text descriptions.
\end{itemize}

For both datasets, to better simulate real-world sequential recommendation scenarios, and prevent data leakage~\cite{data1}, we divide each dataset into 10 periods based on the timestamp of the interactions. 
Subsequently, we split the periods of datasets into training, validation, and testing sets using a ratio of 8:1:1.
To save computing resources, we adopt the same approach as BIGRec, wherein we sample 65536 instances for training without altering the distribution of the original training dataset.
Specifically, for the Steam dataset, we discovered a pronounced imbalance in the distribution of game genres in terms of user interactions. This imbalance results in the system primarily recommending games from these specific genres, which has a minimal impact on the overall metrics.
Therefore, to minimize the Training Burden Without Altering the Conclusions,  we made two adjustments. First, we removed games from the dataset whose respective genres had less than 10,000 interactions overall. This helped us focus on genres with substantial user engagement. Second, we set a maximum limit of 10 user interactions for both datasets to conduct our experiments and perform a thorough analysis.
The final statistical information of the dataset is available in Appendix~\ref{app:4}.




To undertake a thorough investigation into the issue of item-side fairness in LRS, we have implemented two distinct categorizations for partitioning the items in our dataset. Consequently, we will analyze the extent of bias in the LRS using each of these partitioning schemes separately. This approach allows us to conduct a more comprehensive examination of the issue and gain deeper insights into the fairness dynamics within the system.
\begin{itemize}[leftmargin=*]
    \item \textbf{Popularity:}We analyzed the user interaction history by tallying the number of interactions for each item. Subsequently, we ranked the items based on their respective interaction counts. To ensure equitable distribution, we divided these items into five groups, with an equal number of items in each group.
    \item \textbf{Genre:} We extract the genre assigned to each item in the dataset and categorize them accordingly.
\end{itemize}


\subsubsection{Compared Method}
To thoroughly investigate IF in LRS, apart from our implementation of LRS in Section~\ref{sec:bigrec}, we have developed and implemented a typical traditional recommendation system for comparison.
In detail, we select \textbf{SASRec} which is commonly used in previous analyses of IF in the field of recommendation~\cite{SASRec}.
It employs a self-attention mechanism, specifically utilizing a left-to-right attention mechanism. This allows the model to acquire knowledge of sequential patterns and make accurate predictions about subsequent items.


\subsection{Performance on Item-side Fairness}

According to the aforementioned partitioning methods for items in the dataset, we have subdivided the \textbf{RQ2} into two distinct aspects: does LRS provide fairer recommendation results for different groups based on item popularity and genre compared to traditional recommendation systems?

\subsubsection{Popularity Division}
\label{sec:diverse_pop_groups}

First, we conducted a comparative analysis of item popularity distributions between the recommendation results of two recommendation models and the item popularity distribution observed in the user interaction history on the test machine. Our focus was to highlight the discrepancies between these two distributions. The results are presented in Table~\ref{tab:popularity_result}. The main findings are as follows:
\begin{itemize}[leftmargin=*]
\item  Table~\ref{tab:popularity_result} reveals that the fairness performance of the LRS-based model (BIGRec) is significantly inferior to that of the traditional SASRec model on the steam dataset and when considering the top recommendations for the MovieLens1M dataset, BIGRec also shows worse IF performance than SASRec.
\item The recommendation model based on LLM demonstrates a higher degree of variation in fairness metrics compared to traditional models when considering different values of $K$. We attribute this phenomenon to the second step of grounding in BIGRec. In Section~\ref{sec:4_3} of our research, we delve into this issue and conduct a comprehensive investigation to gain a deeper understanding of its implications.
\end{itemize}


\begin{figure}
    \centering
    \begin{subfigure}[b]{\columnwidth}
    \includegraphics[width=0.98\columnwidth]{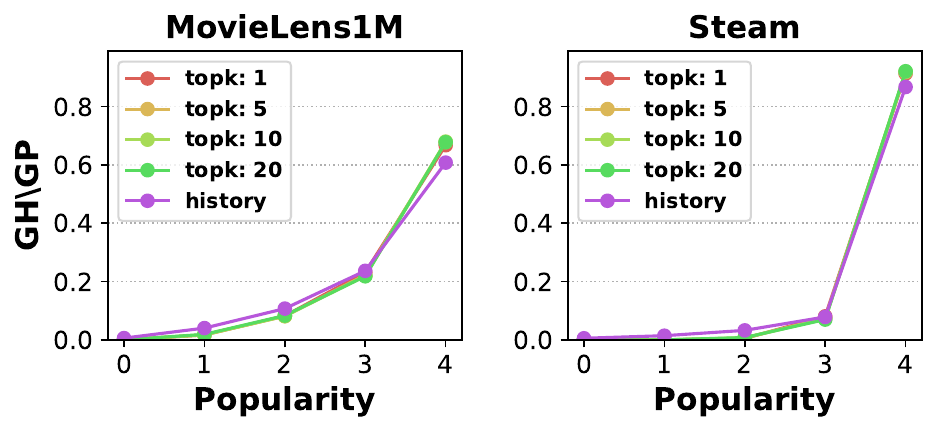}
    \vspace{-0.3cm}

    \caption{SASRec}
    
    \end{subfigure}
    \begin{subfigure}[b]{\columnwidth}
    \includegraphics[width=0.98\columnwidth]{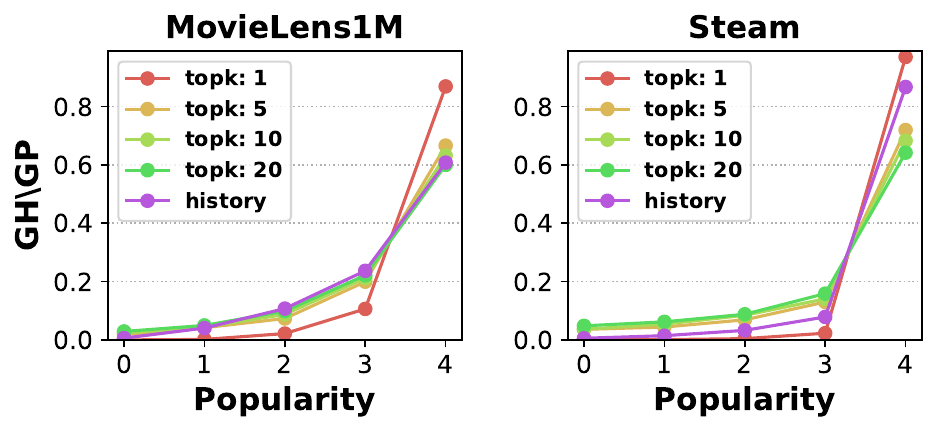}
    \vspace{-0.3cm}

    \caption{BIGRec}
    \end{subfigure}
    \vspace{-0.4cm}

    \caption{
    The proportion distribution of different groups divided by popularity in the top-$K$ recommendation results, compared with the proportion distribution of different groups in the historical interactions (purple curve).}
    \label{fig:proportion_of_pop}
    \vspace{-0.6cm}

\end{figure}

To gain a comprehensive understanding of the unfair issues faced by the popularity group within the LRS, we have analyzed two datasets.
Figure~\ref{fig:proportion_of_pop} visually presents our findings. 
The results indicate that both MovieLens1M and Steam, the top-1 recommendation of LRS excessively recommended group with the highest level of popularity. And the proportions of groups 0 to 3 (representing unpopular items) are consistently lower compared to their historical sequences. 
When considering larger values of $K$, we can observe that this phenomenon is significantly alleviated in both datasets. We attribute this to the efficacy of grounding, and will further discuss this in the subsequent Section~\ref{sec:4_3}.


\subsubsection{Genre Division}

\begin{table*}
\caption{Performance comparison of IF for groups split by genre. The best results are indicated in boldface. We conduct each experiment three times and the average results are reported.}
    \vspace{-0.3cm}

\label{tab:genre_result}
\begin{tabular}{clllllllll}
\hline
Dataset                      & Model  & \textbf{MGU}@1           & \textbf{MGU}@5           & \textbf{MGU}@10          & \textbf{MGU}@20          & \textbf{DGU}@1           & \textbf{DGU}@5           & \textbf{DGU}@10          & \textbf{DGU}@20          \\ \hline
\multirow{2}{*}{MovieLens1M} & SASRec & \textbf{0.0048} & \textbf{0.0037} & \textbf{0.0031} & \textbf{0.0025} & \textbf{0.0224} & \textbf{0.0194} & \textbf{0.0172} & \textbf{0.0152} \\
                             & BIGRec & 0.0068          & 0.0072          & 0.0065          & 0.0060          & 0.0374          & 0.0418          & 0.0382          & 0.0383          \\ \hline
\multirow{2}{*}{Steam}       & SASRec & \textbf{0.0122} & \textbf{0.0105} & 0.0092          & \textbf{0.0075} & \textbf{0.0477} & \textbf{0.0452} & \textbf{0.0442} & 0.0403          \\
                             & BIGRec & 0.0158          & 0.0106          & \textbf{0.0082} & 0.0081          & 0.0487          & 0.0496          & 0.0443          & \textbf{0.0341} \\ \hline
\end{tabular}
\end{table*}

\begin{figure}
    \centering
    \includegraphics[width=0.46\textwidth]{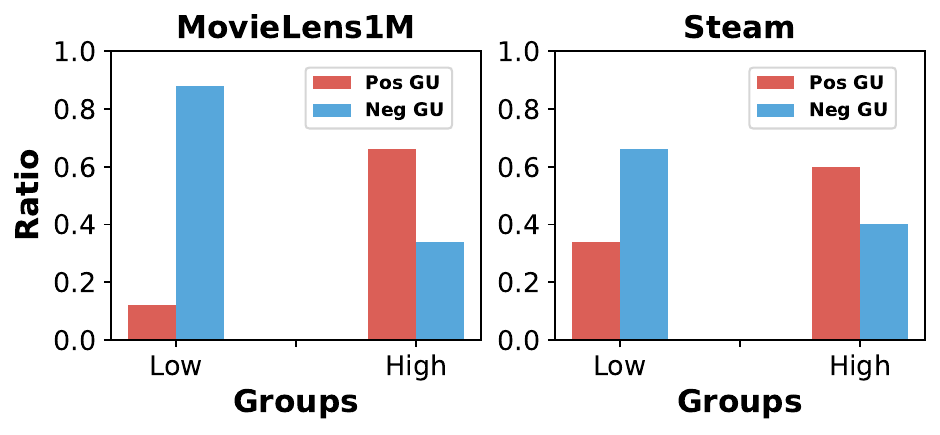}
    \vspace{-0.3cm}
    \caption{Comparison of GU@1 between groups divided by genre.
    We split these genre groups into two parts based on their interaction proportions in historical sequences, and each part has the same number of groups.
    ``Pos GU'' denotes GU@1 > 0, and ``Neg GU'' denotes GU@1 < 0.
    We can observe that high-popularity genres would be over-recommended (Pos GU), while low-popularity genres tend to be overlooked (Neg GU). 
    }
    \label{fig:genre_top1}
    \vspace{-0.5cm}
\end{figure}



%

Subsequently, we evaluate the fairness of SASRec and BIGRec for different genre groups, and the comparison results are reported in Table~\ref{tab:genre_result}.
The main observations are as follows:

\begin{itemize}[leftmargin=*]
    \item The SASRec model achieves superior performance compared to the BIGRec model on almost all metrics. This indicates that, in comparison to conventional models, LRS exhibits significant disparities in fairness within the item genre grouping.

    
    \item In contrast to the inequities observed in popularity, the disparities in item 
    genres are significantly less pronounced. This could be attributed to the larger variety of item genres and their inherently smaller biases. 
\end{itemize}

\subsection{Cause of the Fairness Issues in LRS}
\label{sec:4_3}

In order to enhance our comprehension of the reason why LRS is unfair, we have devised experimental demonstrations in this section to discover its root reason.
Firstly, we focus our attention on the performance of LRS in terms of top-1 recommendation result, which is slightly affected by grounding and shows the fairness issues inherently existing in LLMs themselves \footnote{The reason and details can be found in Appendix~\ref{app:2}.}.
According to Figure~\ref{fig:genre_top1}, we can observe that LRS tends to overly focus on groups with high popularity and excessively recommends items from these groups.
When considering the item genre,  we organized them into groups based on the number of historical interactions. Subsequently, we evenly divided these groups into two parts: "Low" and "High," ensuring that each part contained an equal number of item genres.
The result depicts that similarly to grouping by popularity,  we can observe the LRS's tendency to disproportionately recommend item types that account for a larger proportion of historical interactions. 
Besides, we find that different from the analysis of the popularity group, there are groups with low proportions of group historical interactions that have positive group unfairness.
This means that for genre groups with low frequencies, the model suggests a greater number of items compared to their historical trends.

\begin{figure}[t]
    \centering
    \begin{subfigure}[b]{0.51\columnwidth}
        \centering
        \includegraphics[width=\columnwidth]{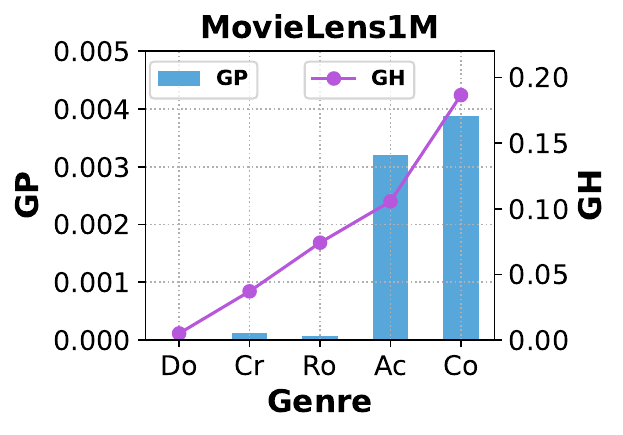}
        \vspace{-0.3cm}
        \caption{Proportions of different groups \wrt genre in top-1 recommendation result, after deleting items of the corresponding genres at training phase, on MovieLens1M data.}
        \label{fig:genre_delete}
    \end{subfigure}%
    \hspace{2mm}
    \begin{subfigure}[b]{0.43\columnwidth}
        \centering
        \includegraphics[width=\columnwidth]{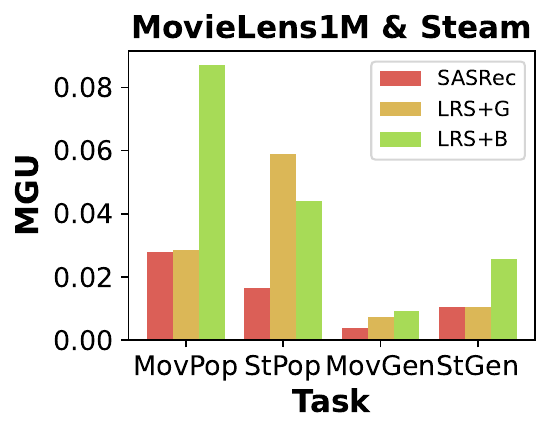}
        \vspace{-0.3cm}
        \caption{Unfairness (MGU@5) comparison among SASRec, LRS with grounding for ranking (LRS + G), and LRS with beamserach for ranking (LRS + B).}
        \label{fig:beamsearch}
    \end{subfigure}
    \vspace{-0.3cm}

    \caption{Notations of (a): \textit{GH} and \textit{GP} denotes the proportions of groups in historical interactions and recommendation results respectively.  The horizontal axis represents movie genres, where ``Do'', ``Cr'', ``Ro'', ``Ac'', and ``Co'' denote \textit{Documentary}, \textit{Crime},  \textit{Romance},  \textit{Action}, and \textit{Comedy} respectively, whose \textit{GH}s increase from left to right.    
    Notations of (b): the horizontal axis represents different tasks, where ``Mov'' and ``St'' denote \underline{Mov}ieLen1M and \underline{St}eam datasets respectively; ``Pop'' and ``Gen'' denote groups divided by \underline{pop}ularity and \underline{gen}re respectively.
    }
    \vspace{-0.6cm}

    \label{fig:length}
\end{figure}

To further explore why that phenomenon occurs across various genre groups, we propose additional experiments to uncover these factors. 
Specifically, we select five genres in MovieLens1M and remove them from our training set.
Then, we verify the fairness metrics of our newly trained LLM on these genre groups as illustrated in Figure~\ref{fig:genre_delete}.
Surprisingly, we found that even after certain item types were removed, there was still a small probability that they would be recommended.
For instance, even if during the supervised fine-tuning stage, models had not encountered movies of the Comedy genre, they could still recommend ``Mighty Ducks, The (1992)'' to users\footnote{More cases and details can be found in Appendix~\ref{app:1}.}. This indicates that during the recommendation process, the models leverage knowledge acquired from their pre-training phase, which potentially affects the fairness of their recommendations.


\begin{figure}
    \centering
    \includegraphics[width=0.48\textwidth]{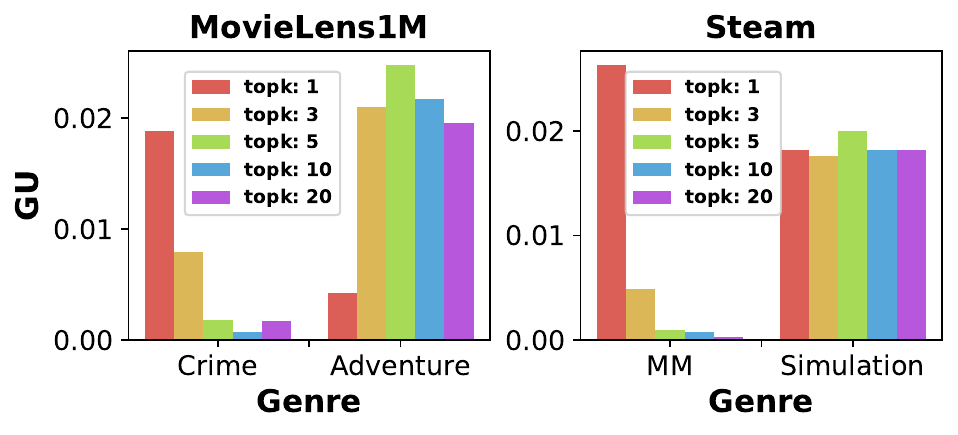}
    \vspace{-0.6cm}

    \caption{The GU (Group Unfairness) of different groups divided by genres in top-$K$ recommendation results.}
    \vspace{-0.5cm}

    \label{fig:special_genre}
\end{figure}

Then, we shall explore what happens as the number of recommended items increases. The primary experimental observations and conclusions are as follows:

\begin{itemize}[leftmargin=*]
    \item  As shown in Figure~\ref{fig:proportion_of_pop}, there is a clear decrease in the proportion of group 4 with the increase of the value of top-$K$, while simultaneously observing an increase in the proportions of groups 0 to 3. Thus we can conclude that the inequity in the popularity of the model decreases as $K$ increases. We attribute this phenomenon to the fact that the original grounding step of BIGRec is not affected by the influence of popularity in specific datasets and consequently recommends a plethora of unpopular items. However, this also results in the prejudice of the model towards low-popularity groups transfer to high-popularity groups.
    \item Besides, according to Figure~\ref{fig:special_genre}, we find that due to the non-uniformity of the embedding space, this step serves to mitigate the inherent unfairness present in certain genres. However, it is important to note that this process may introduce new fairness concerns for other genres or maintain the status. This indicates that the impact of the direct grounding on fairness across different genre groups will vary depending on the distribution of these groups in the embedding space.
    \item Based on our findings, we are considering whether it would be advisable to eliminate the second step of BIGRec and rely solely on beam search to address the issue discussed. To gain a deeper understanding of this matter, we conducted an analysis, as depicted in Figure~\ref{fig:beamsearch}\footnote{Due to the significant cost associated with beam inference, we have specifically compared the circumstances at iteration 5 only.}. The results of our experiments indicate that the use of beam search significantly amplifies inequity, except for the popularity bias observed in the Steam dataset, which can be attributed to the over-correction of BIGRec. This phenomenon is not unique, as it shares similarities with previous discussions in the field of Natural Language Processing. Employing beam search exacerbates various biases inherent in the model~\cite{beamsearch, beamsearch2}.
\end{itemize}

\vspace{-0.3cm}

\subsection{Summary}
Based on the experiments and phenomena mentioned above, we can deduce the following conclusions:
\begin{itemize}[leftmargin=*]
    \item No matter whether observing the grouping of popular items or the grouping of item genres, we have observed unfairness on the item side in LRS. This unfair phenomenon is particularly evident in the top recommendation.
    \item When contemplating the top 1 recommendation, it becomes evident that LRS has a propensity to excessively promote categories with a higher proportion in the training set, and this results in a diminished allocation of recommendations towards items that were originally overlooked. 
    \item LRS can recommend item genres that are never seen in the supervised instruction tuning period, which means the IF in LRS is not only derived from the training set during the fine-tuning phase but also emanates from the semantic priors obtained during the pre-training phase.
\end{itemize}

\section{Methods}
To enhance the IF of LRS, we propose the IFairLRS framework. 
As discussed in Section ~\ref{sec:4_3}, the unfairness stems from two phases: the instruction tuning phase of LLMs (\ie in-learning stage) and the ranking phase of the recommendation system (\ie post-learning stage). 
Accordingly, IFairLRS has two specific strategies: reweighting and reranking strategies to enhance the IF of LRS during the in-learning and post-learning stages, respectively. 
\subsection{Reweighting Strategy}

We first consider reweighting the instruction-tuning samples to reduce the effect of biased training data. Traditional IF methods might add regulation terms~\cite{regulation1, regulation2} or adaptively adjust the sample weights~\cite{ips} during the in-learning stages of conventional recommendation models. However, these methods require the ranking scores over the items of different groups~\cite{regulation1, regulation2} or adaptive calculate the weight of various groups ~\cite{ips}. It is challenging for generative LRS to fulfill these requirements since generative LRS is tuned to maximize the likelihood of the tokens of target item descriptions, instead of calculating the ranking scores over many item candidates like discriminative models.










In light of this, we pre-calculate the sample weights for reweighting before instruction-tuning. Specifically, for the instruction-tuning, we split users' historical interaction sequences $\mathcal{H}$ into the set of training sequences $\mathcal{H}_{tr}$ and the set of target items $\mathcal{H}_{ta}$, where the last item of each sequence is treated as the target item and $(H_i, T_i)$ with $H_i\in \mathcal{H}_{tr}$ and $T_i\in \mathcal{H}_{ta}$ represents the $i$-th instruction-tuning sample. 
If more target items belong to an item group $G$ in $\mathcal{H}_{ta}$, LRS might be tuned biased to this group. 
As shown in~\ref{sec:diverse_pop_groups}, the LLM could learn more bias stemming from unbalanced groups.
As such, we adjust the sample weights of different groups based on their proportions in $\mathcal{H}_{tr}$ and in $\mathcal{H}_{ta}$. 

In detail, we can calculate the interaction proportion $\textbf{GH}(G)$ of a group $G$ in $\mathcal{H}_{tr}$ and $\mathcal{H}_{ta}$ by Equation~\eqref{eq:count}, which are denoted as $\textbf{GH}_{tr}(G)$ and $\textbf{GH}_{ta}(G)$, respectively. 
Given $\textbf{GH}_{tr}(G)$ and $\textbf{GH}_{ta}(G)$, we calculate the weight $W_G$ of group $G$ by:
\begin{equation}
    W_G = \frac{\textbf{GH}_{tr}(G)}{\textbf{GH}_{ta}(G)},
\end{equation}
where $W_G<1$ suggests that there is an amplified proportion of group $G$ in $\mathcal{H}_{ta}$ than that in $\mathcal{H}_{tr}$, and thus we need to reduce the weight of group $G$ in the instruction-tuning stage. 
Moreover, since an item may belong to several groups, the weight $W_i$ of the $i$-th sample $(H_i, T_i)$ can be calculated as: 
\begin{equation}
\label{eq:mean_w}
    W_i = \frac{1}{|\mathcal{G}_i|}\sum_{G \in \mathcal{G}_i} W_G,
\end{equation}
\vspace{-0.2cm}

where $\mathcal{G}_i$ denotes the groups that the target item $T_i$ belongs to.
Finally, we utilize $W_i$ to reweight each sample and the loss of the $i$-th sample $(H_i, T_i)$ is as follows: 
\begin{equation}
    \textbf{L}(H_i, T_i) = W_i * \mathcal{L}(H_i, T_i),
\end{equation}
\vspace{-0.6cm}

where $\mathcal{L}(\cdot, \cdot)$ denotes the instruction-tuning loss of LLMs. 

\vspace{-0.4cm}

\begin{table*}
    \centering
    \caption{The IF of IFairLRS with different strategies for groups split by popularity. The best results of all methods are indicated in boldface. IFairLRS(RW) denotes the reweighting strategy and IFairLRS(RR) denotes the reranking strategy. We conduct each experiment three times and report the average results.
}
    \vspace{-0.4cm}

\setlength{\tabcolsep}{2mm}{
\resizebox{\textwidth}{!}{
\begin{tabular}{c|c|llllll|ll}
\hline
                              &                                   & \multicolumn{6}{c|}{Fairness}                                                                                                                                                                                                                             & \multicolumn{2}{c}{Accuracy}                                                   \\ \cline{3-10} 
\multirow{-2}{*}{Dataset}     & \multirow{-2}{*}{Model}           & \multicolumn{1}{c}{\textbf{MGU}@1 $\downarrow$}               & \multicolumn{1}{c}{\textbf{MGU}@5 $\downarrow$}               & \multicolumn{1}{c}{\textbf{MGU}@20 $\downarrow$}              & \multicolumn{1}{c}{\textbf{DGU}@1 $\downarrow$}               & \multicolumn{1}{c}{\textbf{DGU}@5 $\downarrow$}               & \multicolumn{1}{c|}{\textbf{DGU}@20 $\downarrow$}             & \multicolumn{1}{c}{\textbf{NDCG}@5 $\uparrow$}             & \multicolumn{1}{c}{\textbf{HR}@5 $\uparrow$}               \\ \hline
                              & BIGRec                            & 0.1044                                  & 0.0285                                  & 0.0128                                  & 0.3917                                  & 0.0956                                  & 0.0415                                  & \textbf{0.0299}                                  & 0.0345                                  \\
                              & \cellcolor[HTML]{EFEFEF}IFairLRS(RW) & \cellcolor[HTML]{EFEFEF}\textbf{0.0990} & \cellcolor[HTML]{EFEFEF}\textbf{0.0198} & \cellcolor[HTML]{EFEFEF}0.0160          & \cellcolor[HTML]{EFEFEF}\textbf{0.3585} & \cellcolor[HTML]{EFEFEF}\textbf{0.0527} & \cellcolor[HTML]{EFEFEF}0.0544          & \cellcolor[HTML]{EFEFEF}0.0295          & \cellcolor[HTML]{EFEFEF}0.0341          \\
\multirow{-3}{*}{MovieLens1M} & \cellcolor[HTML]{EFEFEF}IFairLRS(RR) & \cellcolor[HTML]{EFEFEF}0.1074          & \cellcolor[HTML]{EFEFEF}0.0352          & \cellcolor[HTML]{EFEFEF}\textbf{0.0125} & \cellcolor[HTML]{EFEFEF}0.4011          & \cellcolor[HTML]{EFEFEF}0.1190          & \cellcolor[HTML]{EFEFEF}\textbf{0.0413} & \cellcolor[HTML]{EFEFEF}0.0297 & \cellcolor[HTML]{EFEFEF}\textbf{0.0348} \\ \hline
                              & BIGRec                            & 0.0411                                  & 0.0590                                  & 0.0900                                  & 0.1586                                  & 0.1987                                  & 0.3050                                  & 0.0311                                  & 0.0346                                  \\
                              & \cellcolor[HTML]{EFEFEF}IFairLRS(RW) & \cellcolor[HTML]{EFEFEF}\textbf{0.0371} & \cellcolor[HTML]{EFEFEF}0.0691          & \cellcolor[HTML]{EFEFEF}0.0959          & \cellcolor[HTML]{EFEFEF}\textbf{0.1410} & \cellcolor[HTML]{EFEFEF}0.2290          & \cellcolor[HTML]{EFEFEF}0.3208          & \cellcolor[HTML]{EFEFEF}0.0311          & \cellcolor[HTML]{EFEFEF}0.0341          \\
\multirow{-3}{*}{Steam}       & \cellcolor[HTML]{EFEFEF}IFairLRS(RR) & \cellcolor[HTML]{EFEFEF}0.0417          & \cellcolor[HTML]{EFEFEF}\textbf{0.0179} & \cellcolor[HTML]{EFEFEF}\textbf{0.0377} & \cellcolor[HTML]{EFEFEF}0.1610          & \cellcolor[HTML]{EFEFEF}\textbf{0.0774} & \cellcolor[HTML]{EFEFEF}\textbf{0.1308} & \cellcolor[HTML]{EFEFEF}\textbf{0.0312} & \cellcolor[HTML]{EFEFEF}\textbf{0.0348} \\ \hline
\end{tabular}
}}
    \label{tab:modules_popylarity}
    \vspace{-0.4cm}

\end{table*}

\subsection{Reranking Strategy}
In addition to the reweighting strategy for the in-learning stage, we consider reranking the items in the post-learning stage. Given the top-$K$ recommendations from LRS, the reranking strategy can calculate \textbf{GU}$(G)@K$ to measure the unfairness of each group $G$. To enhance IF, we can rerank the items of different groups by revising \textbf{GU}$(G)@K$ as a punishment term.

Specifically, we calculate $\textbf{GU}(G)@K$ of each group $G$ \wrt varying $K$ values and then aggregate them to obtain the punishment term $\textbf{U}_G$ for group $G$. We vary $K$ to calculate $\textbf{U}_G$ because the grounding strategy of LRS might over-correct the bias when $K$ is large (as shown in Section~\ref{sec:4_3}). By aggregating $\textbf{GU}(G)@K$ for punishment, we can better regulate the unfairness across different top-$K$ recommendations. 
Formally, we have 
\begin{equation}
\label{eq:UG}
\begin{aligned}
    \textbf{U}_G = \sum_{K \in \mathcal{K}}\gamma_K*\textbf{GU}(G)@K, 
\end{aligned}
\end{equation}
\vspace{-0.4cm}

where $\mathcal{K}$ includes the possible values of $K$ and $\gamma_K$ adjusts the weights of unfairness with different top-$K$ values. 
In this work, we use $\gamma_K = {K} / {\sum_{K^{'} \in \mathcal{K}} K^{'}}$, where $\gamma_K$ increases with $K$ increasing. In this way, it pays more attention to alleviating the over-correction of the grounding strategy of LRS when $K$ is large. 
Notably, we further normalize the $\textbf{U}(G)$ into $[-1, 1]$ by \begin{equation}
    \hat{\textbf{U}}_G = \frac{\textbf{U}_G}{\max \{\left|\textbf{U}_{G^{'}}\right|\}_{G^{'}\in\mathcal{G}}},
\end{equation}
\vspace{-0.2cm}

where $\left|\textbf{U}_{G^{'}}\right|$ takes the absolute value of $\textbf{U}_{G'}$. 
Since an item may belong to several groups, we calculate the punishment $\hat{\textbf{U}}_i$ of the $i$-th item by:
\begin{equation}
    \hat{\textbf{U}}_i = \frac{1}{|\mathcal{G}_i|}\sum_{G \in \mathcal{G}_i} \hat{\textbf{U}}_G,
\end{equation}
\vspace{-0.2cm}

where $\mathcal{G}_i$ denote the groups that the $i$-th item belong to.
Eventually, we add $\hat{\textbf{U}}_i$ to Equation~\eqref{eq:distance} for item reranking:
\begin{equation}
\label{eq:rerank_distance}
        \Tilde{D}_i = \frac{D_i}{(1-\hat{\textbf{U}}_i)^\alpha},
\end{equation}

where a hyperparameter $\alpha>0$ is to regulate the influence of punishment, and $D_i$ denotes the L2 distance between the embedding of the $i$-th item and the embedding of the generated item description by LLMs. Intuitively, the items with positive $\hat{\textbf{U}}_i$ are over-recommended (see explanation in Equation~\ref{eq:GU}) and thus they get a larger L2 distance $\Tilde{D}_i$ in Equation~\ref{eq:rerank_distance} for lower recommendation probabilities. In contrast, the items with negative $\hat{\textbf{U}}_i$ will have higher recommendation probabilities with smaller $\Tilde{D}_i$. 

\vspace{-0.2cm}

\section{Experiment}
In this section, we conduct experiments on the two datasets to answer the following research question:

\begin{itemize}[leftmargin=*]
    \item \textbf{RQ3:} Can IFairLRS with two strategies effectively enhance the IF of LRS?
\end{itemize}

\subsection{Experiment Setting}
Following Section~\ref{sec:4_1_1}, we conduct experiments on two datasets: MovieLens1M and Steam. The dataset statistics and division can be found in Section~\ref{sec:4_1_1}. 
In this section, we not only evaluate the fairness but also evaluate the accuracy of recommendation models, so as to examine whether IFairLRS will damage the recommendation accuracy of LRS while pursuing IF. 
Following the previous work~\cite{dros, bigrec}, we use two commonly used evaluation metrics for accuracy measurements: Normalized Discounted Cumulative Gain (\textbf{NDCG}) and Hit Ratio (\textbf{HR}). 
Both metrics are computed under the all-ranking protocol~\cite{dros}. 

We implement all of the methods using PyTorch. We follow the setting of BIGRec using LLaMA~\cite{LLaMA} as the base LLM for instruction-tuning. We use the Adam~\cite{Adam} as the optimizer with a learning rate of $1e-3$, a batch size of 128, and we tune the weight decay in the range of [$1e-2$, $1e-3$, $1e-4$, $1e-5$, $1e-6$].
For the reranking strategy of IFairLRS, we tune its hyperparameter $\alpha$ in Equation~\eqref{eq:rerank_distance} in the range of [0,0.1] with step 0.01. We utilize the validation sets to calculate \textbf{GU}(G) for Equation~\ref{eq:UG} because it includes user interactions temporally closer to the testing sets.

\subsection{Performance Comparison}
To answer \textbf{RQ3}, we compare the fairness and accuracy of BIGRec with the reweighting and reranking strategies of IFairLRS. 
Specifically, we divide item groups in two ways: popularity and genre, and then measure the recommendation fairness and accuracy in terms of $\textbf{MGU}@K$, $\textbf{DGU}@K$, $\textbf{NDCG}@K$, and $\textbf{HR}@K$.


\begin{table*}
\centering
\caption{The IF of IFairLRS with different strategies for groups split by genre. The best results of all methods are indicated in boldface. IFairLRS(RW) denotes the reweighting strategy and IFairLRS(RR) represents the reranking strategy. We conduct each experiment three times and report the average results.
}
\vspace{-0.4cm}

\setlength{\tabcolsep}{2mm}{
\resizebox{\textwidth}{!}{
\begin{tabular}{c|c|llllll|ll}
\hline
                              &                                   & \multicolumn{6}{c|}{Fairness}                                                                                                                                                                                                                             & \multicolumn{2}{c}{Accuracy}                                                   \\ \cline{3-10} 
\multirow{-2}{*}{Dataset}     & \multirow{-2}{*}{Model}           & \multicolumn{1}{c}{\textbf{MGU}@1 $\downarrow$}               & \multicolumn{1}{c}{\textbf{MGU}@5 $\downarrow$}               & \multicolumn{1}{c}{\textbf{MGU}@20 $\downarrow$}              & \multicolumn{1}{c}{\textbf{DGU}@1 $\downarrow$}               & \multicolumn{1}{c}{\textbf{DGU}@5 $\downarrow$}               & \multicolumn{1}{c|}{\textbf{DGU}@20 $\downarrow$}             & \multicolumn{1}{c}{\textbf{NDCG}@5 $\uparrow$}             & \multicolumn{1}{c}{\textbf{HR}@5 $\uparrow$}               \\ \hline
                              & BIGRec                            & 0.0068                                  & 0.0072                                  & 0.0060                                  & 0.0374                                  & 0.0418                                  & 0.0383                                  & \textbf{0.0299}                         & \textbf{0.0345}                         \\
                              & \cellcolor[HTML]{EFEFEF}IFairLRS(RW) & \cellcolor[HTML]{EFEFEF}\textbf{0.0050} & \cellcolor[HTML]{EFEFEF}0.0067          & \cellcolor[HTML]{EFEFEF}0.0056          & \cellcolor[HTML]{EFEFEF}\textbf{0.0268} & \cellcolor[HTML]{EFEFEF}0.0359          & \cellcolor[HTML]{EFEFEF}0.0366          & \cellcolor[HTML]{EFEFEF}0.0291          & \cellcolor[HTML]{EFEFEF}0.0337          \\
\multirow{-3}{*}{MovieLens1M} & \cellcolor[HTML]{EFEFEF}IFairLRS(RR) & \cellcolor[HTML]{EFEFEF}0.0089          & \cellcolor[HTML]{EFEFEF}\textbf{0.0054} & \cellcolor[HTML]{EFEFEF}\textbf{0.0036} & \cellcolor[HTML]{EFEFEF}0.0439          & \cellcolor[HTML]{EFEFEF}\textbf{0.0339} & \cellcolor[HTML]{EFEFEF}\textbf{0.0184} & \cellcolor[HTML]{EFEFEF}0.0291          & \cellcolor[HTML]{EFEFEF}0.0336          \\ \hline
                              & BIGRec                            & 0.0158                                  & 0.0106                                  & 0.0081                                  & 0.0487                                  & 0.0496                                  & 0.0341                                  & 0.0311                                  & 0.0346                                  \\
                              & \cellcolor[HTML]{EFEFEF}IFairLRS(RW) & \cellcolor[HTML]{EFEFEF}\textbf{0.0138} & \cellcolor[HTML]{EFEFEF}0.0092          & \cellcolor[HTML]{EFEFEF}0.0078          & \cellcolor[HTML]{EFEFEF}\textbf{0.0483} & \cellcolor[HTML]{EFEFEF}0.0440          & \cellcolor[HTML]{EFEFEF}0.0331          & \cellcolor[HTML]{EFEFEF}\textbf{0.0316} & \cellcolor[HTML]{EFEFEF}\textbf{0.0348} \\
\multirow{-3}{*}{Steam}       & \cellcolor[HTML]{EFEFEF}IFairLRS(RR) & \cellcolor[HTML]{EFEFEF}0.0184          & \cellcolor[HTML]{EFEFEF}\textbf{0.0084} & \cellcolor[HTML]{EFEFEF}\textbf{0.0065} & \cellcolor[HTML]{EFEFEF}0.0541          & \cellcolor[HTML]{EFEFEF}\textbf{0.0396} & \cellcolor[HTML]{EFEFEF}\textbf{0.0284} & \cellcolor[HTML]{EFEFEF}0.0311          & \cellcolor[HTML]{EFEFEF}0.0347          \\ \hline
\end{tabular}
}}
\vspace{-0.3cm}

\label{tab:modules_genre}
\end{table*}

\subsubsection{Popularity Division}
To validate the effectiveness of IFairLRS on diverse popularity groups, we compare BIGRec with the reweighting strategy and reranking strategy of IFairLRS. 
The fairness and accuracy comparison is reported in Table~\ref{tab:modules_popylarity}, from which the main observations are as follows:
\begin{itemize}[leftmargin=*]
    \item 
    The reweighting strategy in IFairLRS proves effective in enhancing the IF of LRS. On the MovieLens1M dataset, this strategy yields the best $\textbf{MGU}@1$ and $\textbf{DGU}@1$, showcasing improvements of 5\% and 8.4\% over BIGRec, respectively. Similarly, on the Steam dataset, the reweighting strategy demonstrates fairness improvements of 9.7\% and 11.1\% on $\textbf{MGU}@1$ and $\textbf{DGU}@1$, respectively. These findings indicate that the reweighting strategy successfully promotes the calibration between top-1 recommendations and users' historical interactions across item groups.  
    \item 
    The reranking strategy reveals better fairness when $K$ is large while its effectiveness is limited with small $K$. For instance, IFairLRS(RR) has comparable or slightly worse fairness \wrt $\textbf{MGU}@1$ and $\textbf{DGU}@1$ on two datasets while it shows better performance \wrt $\textbf{MGU}@20$ and $\textbf{DGU}@20$, especially on the Steam dataset. The possible reason is that reranking can better balance the fairness across groups when $K$ is large with more recommendation slots. 

    \item 
    BIGRec and IFairLRS with two strategies demonstrate comparable accuracy in terms of $\textbf{NDCG}@5$ and $\textbf{HR}@5$. The performance difference among BIGRec, IFairLRS(RW), and IFairLRS(RR) for $\textbf{NDCG}@5$ on both datasets fluctuates within 1\%. Similarly, for $\textbf{HR}@5$, the deviation remains within 5\%. These results validate that IFairLRS(RW) and IFairLRS(RR) do not enhance IF at the significant expense of recommendation accuracy. 

    \item As the value of top-$K$ increases, the observed changes in metrics for IFairLRS(RW) remain consistent with the trend of BIGRec.
    Even on the $\textbf{MGU}@20$ and $\textbf{DGU}@20$, the IFairLRS with the reweighting strategy obtains worse fairness for various popularity groups.
    Hence, the issues that exist in the grounding are not solved by the reweighting strategy.

    \item The IFairLRS(RR) obtains comparable or slightly worse $\textbf{MGU}@1$ and $\textbf{DGU}@1$. We analyze the reasons for this phenomenon. 
    For instance, as shown in Figure~\ref{fig:proportion_of_pop}, the $\textbf{GU}@K$ of group 4 reduces from positive to negative with the increase of the value of top-$K$ on the Steam dataset.
    We vary $K$ to calculate the punishment of the group and the $\textbf{GU}@K$ with higher $K$ has more weight.
    Hence, the punishment of group 4 is the opposite of $GU@1$, which leads to the IFairLRS(RR) obtaining comparable or slightly worse $\textbf{MGU}@1$ and $\textbf{DGU}@1$.
\end{itemize}
\vspace{-0.3cm}

\subsubsection{Genre Division}
To validate the effectiveness of the reweighting strategy and the reranking strategy of IFairLRS on different genre groups, we conduct experiments to compare their performance. 
The fairness and accuracy performance is reported in Table~\ref{tab:modules_genre}. 
In the table, we have the observations as follows:

\begin{itemize}[leftmargin=*] 
    \item Both strategies exhibit the effectiveness of improving the fairness of LRS. 
    On MovieLens1M, the reweighting strategy enhances 0.0018 for $\textbf{MGU}@1$ and 0.0106 for $DGU@1$ over BIGRec (as much as 30\% and 28.3\%, respectively). 
    On Steam, IFairLRS with the reweighting strategy obtains 12.7\% and 11.1\% improvement on $\textbf{MGU}@1$ and $\textbf{DGU}@1$, respectively.
    Furthermore, on MovieLens1M, IFairLRS with the reranking strategy obtains the best $\textbf{MGU}@20$ and $\textbf{DGU}@20$, which outperform the BIGRec 40\% and 52\%, respectively.  
    The $\textbf{MGU}@20$ and the $\textbf{DGU}@20$ of IFairLRS with the reranking strategy are 0.0016 and 0.0057 lower than the BIGRec (as much as 19.7\% and 16.7\%) on Steam. 
    These suggest that the reranking strategy could also mitigate the unfairness across various genre groups when $K$ is large. 
    
    \item Both strategies improve the fairness of LRS without compromising the recommendation accuracy.
    On MovieLens1M, two strategies have the accuracy decrease by up to 0.001 \wrt $\textbf{NDCG}@5$ and 0.0012 \wrt $\textbf{HR}@5$.
    On Steam, $\textbf{NDCG}@5$ and $\textbf{HR}@5$ fluctuate within 3\%. This validates that although IFairLRS adjusts the recommendation proportion of different item groups for fairness, it does not reduce the number of recommended positive items (\ie users' liked items). 
\end{itemize}

\vspace{-0.4cm}
\section{Conclusion}
\vspace{-0.1cm}

In this work, we examined the item-side unfairness of LRS and compared it with that of conventional recommendation models. 
We found that LRS is not only significantly influenced by the popularity factor but also affected by the inherent semantic biases within LLMs. 
These findings highlight the necessity of improving the IF of LRS. 
To achieve this goal, we conducted a preliminary exploration by proposing a concise and effective framework called IFairLRS.
IFairLRS adapts the reweighting and reranking strategies from traditional IF methods to the in-learning and post-learning stages of LRS, respectively. Specifically, the reweighting strategy adjusts the weights of training samples to reduce the effect of bias and the reranking strategy reranks the item candidates by introducing a punishment term. 
We conducted extensive experiments on two real-world datasets, MovieLens1M and Steam, demonstrating the effectiveness of our proposed IFairLRS in improving the IF of LRS without sacrificing the recommendation accuracy.

We firmly believe that enhancing the IF of LRS is vitally important for improving the trustworthiness of LRS and contributing to making the Web for good. 
IFairLRS with two strategies presents a preliminary exploration, and it is promising to design more effective fairness-oriented methods specifically tailored for LRS in future work. 
Moreover, our analysis has found that the pre-training of LLMs affects the IF of LRS while it is challenging to quantify and alleviate such impact, leaving ample room for future research. 
Lastly, we mainly investigate the IF of LRS based on the group division of popularity and genre. Future efforts might shed light on the fairness of more group divisions (\eg different item uploader groups) and LLMs, individual-level fairness, and long-term fairness.

\begin{acks}
This work is supported by the National Key Research and Development Program of China (2022YFB3104701), the National Natural Science Foundation of China (U21B2026 and 62272437), and the CCCD Key Lab of Ministry of Culture and Tourism.
\end{acks}

\bibliographystyle{ACM-Reference-Format}
\balance
\bibliography{reference}

\appendix
\section{Datasets details}
\label{app:4}

\begin{table}[h]
    \caption{Statistics of the experimental datasets.}
    \vspace{-0.3cm}

    \label{tab:datasets}
    \begin{tabular}{cccc}
                \hline 
    Datasets     & \#Items & \#Interactions & \#Sequences \\
                \hline
    MovieLens1M & 3,883  & 1,000,209      & 939,809   \\
    Steam       & 14,662 & 7,793,036      & 1,620,946 
                \\\hline    
\end{tabular}  
\end{table}
The statistics of the experimental datasets can be seen in Table~\ref{tab:datasets}.

\section{Relationship between outputs of LLM and top 1 recommendations}
\label{app:2}

\begin{table}[h]
\centering
\caption{The number of outputs of LLM existing in the datasets.}
\begin{tabular}{lcc}
\hline
   Existance & \multicolumn{1}{l}{MovieLens1M} & \multicolumn{1}{l}{Steam} \\ \hline
Yes  & 93667                     & 160696                    \\
No & 314                       & 1399                      \\ \hline
\end{tabular}
\label{tab:exist}
\end{table}
As shown in Table~\ref{tab:exist}, we find that most of the outputs generated by the LLM exist in the datasets.
Because we use the $L2$ distance to ground the outputs from the recommendation space to the actual item space.
So, the grounding step does not modify these outputs existing in the dataset in the top-1 recommendation system.
The impact of grounding for the top-1 recommendation results is very slight.
Hence we could utilize the results of the top-1 recommendations to represent the performance of the LLM.

\section{Analysis of Deleting Genre on MovieLens1M}
\label{app:1}
We delete the items belonging to ``Comedy'' both in the historical interactions and target items in the training set.
We find that the LRS continues to recommend movies in ``Comedy''.
The outputs of LLM are two folds:
\begin{itemize}[leftmargin=*]
    \item The output of LLM is in the dataset. For example, the LLM generates the movie ``Airplane! (1980)'', which is a comedy movie and it is in the dataset.
    \item The output of LLM is not in the dataset. For instance, the LLM generates the movie ``Mighty Ducks, The (1992)'', which is not in the dataset, and the nearest movie in the embedding space is ``The Mighty Ducks (1992)''.
\end{itemize}

\section{Analysis of Difference between MovieLens1M and Steam}

\begin{figure}[h]
    \centering
    \includegraphics[width=0.3\textwidth]{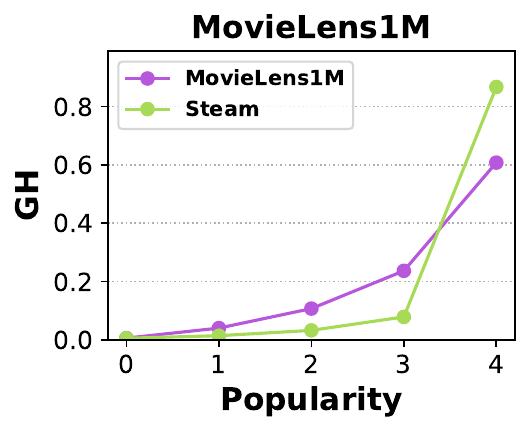}
    \caption{The proportions of various popularity groups of MovieLens1M and Steam.
    }
    \label{fig:app3}
\end{figure}

As shown in Figure~\ref{fig:app3}, we can observe that Steam has a larger popularity bias compared to MovieLens1M.
The proportion of the most popular group in Steam is significantly higher than in MovieLens1M.

\end{document}